\documentclass[epsfig,12pt]{article}
\usepackage{epsfig}
\usepackage{graphicx}
\usepackage{array}
\usepackage{color}
\usepackage{bm}
\usepackage{cite}
\usepackage{geometry}
\usepackage{latexsym}
 \usepackage{amsmath, amssymb, amscd, xypic, graphicx}

\newcommand{\beq}{\begin{equation}}   
\newcommand{\eeq}{\end{equation}}
\newcommand{\beqn}{\begin{eqnarray}}   
\newcommand{\eeqn}{\end{eqnarray}}

\newcommand{\pt}{\partial}

\newcommand*\xbar[1]{%
 \kern0.5ex%
  \hbox{%
   \kern0.2ex%
      \vbox{%
      \hrule height 0.5pt % The actual bar
      \kern0.5ex%         % Distance between bar and symbol
      \hbox{%
        \kern-0.1em%      % Shortening on the left side
        \ensuremath{#1}%
        \kern-0.1em%      % Shortening on the right side
      }%
    }%
  }%
}

\newcommand{\gsim}{\lower.7ex\hbox{$
\;\stackrel{\textstyle>}{\sim}\;$}}
\newcommand{\lsim}{\lower.7ex\hbox{$
\;\stackrel{\textstyle<}{\sim}\;$}}
\begin{document}

\begin{titlepage}
\begin{flushright}
Preprint ITEP-146\\
September 1985
\end{flushright}

\vspace{5mm}

\begin{center}
{  \Large \bf  
Operator Product Expansion and Calculation \\[2mm]

of the Two-Loop Gell-Mann-Low  Function}
 
 \vspace{6mm}

{\Large A. I. Vainshtein and M. A. Shifman} 

\vspace{3mm}

{\it  Institute of Theoretical and Experimental Physics, State Commission for the Utilization of Atomic Energy 

Yad. Fiz. {\bf 44}, 498-506 (August 1986)}

(Submitted 25 September 1985)

 English Translation: 

Sov. J. Nucl. Phys., {\bf 44} (2), 321-325, 1986 
\end{center}

\vspace{4mm}

\vspace{6mm}

\begin{center}
{\large\bf Abstract}
\end{center}

A simple method is developed that makes it possible to determine the $k$-loop coefficient of the
$\beta$-function if the operator	product expansion for certain polarization operators in the $(k -1)$
loop is known. The calculation of the two-loop coefficient of the Gell-Mann-Low function becomes trivial -- it reduces to a few algebraic operations on already known expressions. As examples, spinor, scalar, and supersymmetric electrodynamics are considered. Although the
respective results for  $\beta^{(2)}$ are known in the literature, both the method of calculation and
certain points pertaining to the construction of the operator product expansion are new.

\end{titlepage}

\section{Introduction}

The present paper is devoted to the calculation of the
calculation of the charge renormalization in gauge theories. Although the history of the problem extends over decades, we should like to propose a method for the determination of $\beta$-functions which, in our view, is of interest both for its simplicity and for certain other merits. The main idea is as follows. By separating out the integration over one of the virtual lines in the graphs for the effective Lagrangian in the external field method, we reduce the problem of the 
$k$-loop $\beta$-function to the construction of an operator product expansion in the $(k - 1)$ loop. More precisely, since we are studying the effective charge, in the operator product expansion we need only one term, proportional to the square of the field strength tensor of  the external field. The calculation of $\beta^{(2)}$ (the second coefficient in the Gell-Mann-Low function) becomes extremely simple. If we make use of certain already known results, the determination of 
$\beta^{(2)}$ reduces to a few purely algebraic operations that, in essence, do not require even a single integration.

The resulting expression for the effective action has from the outset a one-logarithm form. Therefore, accuracy in the ultraviolet-regularization procedure is not required. 

At present, dimensional regularization (dimensional reduction in supersymmetric theories) is most often used for this purpose. The proposed method makes it possible to work directly in four-dimensional space.

As examples we find $\beta^{(2)}$ in spinor, scalar, and supersymmetric electrodynamics. The treatment of the latter two cases is also of interest in that new points in such a well studied problem as the construction of the operator product expansion (OPE) are rather unexpectedly revealed.

We start from the operator product expansion for the polarization operator
\beq
\Pi_{\mu\nu}(k) = i\int e^{ikx}d^4x \langle T\{ J_\mu(x) J_\nu (0)\}\rangle\,,\nonumber
\eeq
where$J_\mu (x)$ is the electromagnetic current of the scalar particles. It is found that in the one-loop approximation the coefficient of $F_{\mu\nu}^{\,\,2}$  is determined not only by the contribution of virtual momenta $p\sim k$ but also by the region of momenta $p$
of the order of the momenta of the external field $F_{\mu\nu}$. Nevertheless
the result obtained in this way is the correct result for the OPE coefficient, since a change in the momentum of the external field begins to affect $C_n$ only on the scale $\sim k$.
\begin{flushright}
\color{blue} 321
\end{flushright}

\section{General elements}

%\vspace{-2mm}

First of all we shall formulate certain points that apply in equal measure to all the problems that will be discussed below. We introduce in this section all the necessary notation and explain the general strategy. The initial Lagrangian has the form
\beq
{\mathcal L} = - \frac{1}{4e_0^2} F_{\mu\nu} F^{\mu\nu} +{\rm matter}
\label{1}
\eeq
where $e_0$ is the bare coupling constant (charge), $F_{\mu\nu} $ is the photon-field strength tensor,
\beq
F_{\mu\nu} = \partial_\mu A_\nu -\partial_\nu A_\mu\nonumber
\eeq
and  ``matter" in (\ref{1}) correspond to the usual kinetic terms of the matter fields (in the supersymmetric model a Yukawa coupling is added too; see below). The fields in (\ref{1})  are bare (unrenormalized) fields, i.e., fields normalized at the ultraviolet cutoff mass $M_0$.

Next, starting from the Lagrangian (\ref{1}), we calculate the effective Lagrangian that takes account of virtual fluctuations with momenta $p$,
\beq
\mu\leq p\leq M_0
\eeq
where $\mu$ is a running parameter -- the so-called normalization point:
\beq
{\mathcal L}_{\rm eff} = - \frac{1}{4e^2(\mu)} F_{\mu\nu} F^{\mu\nu} +{\rm other\,\, structures}\,.
\label{3}
\eeq
	The tensor $F_{\mu\nu} $ in (\ref{3}) must be regarded as the external field, and the coefficient multiplying $F_{\mu\nu}^{\,\,2}$ contains the effective (i.e., normalized at the point $\mu$) charge. The Lagrangian ${\mathcal L}_{\rm eff}$ is a fully appropriate Lagrangian in respect of fluctuations with frequencies smaller than $\mu$. After this definition of the charge, the reader familiar with the background-field (external-field) method [2] will understand immediately that it is in the framework of this method that we intend to work.

It is obvious that $e^2(\mu) $ depends on $e_0$ and $M_0/\mu$. The Gell- Mann-Low function $\beta (\alpha)$  is defined as
\beq
\beta(\alpha) = \left. \frac{\partial  \alpha(\mu)}{\partial \log\mu} \right|_{\alpha_0,\,M_0\,{\rm fixed}}
\eeq
where $\alpha =  e^2/4\pi$.

In the two-loop approximation
\begin{flushright}
\color{blue} 321
\end{flushright}
\beq
\frac{1}{\alpha}= \frac{1}{\alpha_0}
+\beta^{(1)}\log\frac{M_0}{\mu} +\beta^{(2)}\alpha_0 \log\frac{M_0}{\mu} +{\rm terms\,\, not\,\, containing} \,\, M_0\,,
\eeq
and, consequently,
\beq
\beta(\alpha ) = \alpha^2 \left(\beta^{(1)} + \alpha \beta^{(2)} + ...\right)
\eeq
The coefficient $\beta^{(1)}$ is determined by the simplest, one-loop graph and, of course, does not require any commentary (see, e.g., Ref. 1; the results are collected in the Table). Our problem is to find $\beta^{(2)}$  in the simplest possible way.

\begin{table}[h!]
\begin{center}
\caption{\small \em Coefficients of the Gell-Mann--Low function (defined in (5) and (6)).}
\label{tab1}
\begin{tabular}{c|c|c|c}
\hline
\hline
 & {\small Spinor electrodyn.} $\rule{0mm}{6mm}$& {\small Scalar  electrodyn.} & {\small Supersymmetric electrodyn.}\\[2mm]
 \hline
$ \beta^{(1)}\rule{0mm}{6mm}$  & $\frac{2}{3\pi} $ & $ \frac{1}{6\pi}$ & $\frac{1}{\pi}$ \\[3mm]
$ \beta^{(2)}$&$\frac{1}{2\pi^2} $&$ \frac{1}{2\pi^2}$& $\frac{1}{\pi^2}$
\end{tabular}
\end{center}
\end{table}

\section{Spinor electrodynamics}

The model includes the photon and electron and is described by the Lagrangian
\beq
{\mathcal L} = - \frac{1}{4e_0^2} F_{\mu\nu} F^{\mu\nu} + \bar\psi i\gamma^\mu D_\mu\psi\,,
\eeq
where $\psi$  is the Dirac spinor,
\beq
iD_\mu = i\partial_\mu + A_\mu\,,
\eeq
and the mass term has been omitted. The effective Lagrangian (\ref{3}) in the two-loop approximation is determined by the graph in Fig. 1, where the solid line denotes the electron propagator in the external photon field. In fact, in the expansion of the propagator in the photon field one needs to keep only terms $O(F_{\mu\nu})$  and $O(F_{\mu\nu}^{\,\,2})$, since no other terms lead to an $F_{\mu\nu}^{\,\,2}$ structure in ${\mathcal L}_{\rm eff}$. 

Furthermore, it is obvious that the photon Green function in Fig. 1 corresponds to the propagation of a free photon:
\beq
D_\mu\nu (x) = g_{\mu\nu} \,\frac{ie_0^2}{4\pi^2}\, \frac{1}{x^2}\,. 
\eeq
\begin{flushright}
\color{blue} 322
\end{flushright}
The corresponding expression for the two-loop ${\mathcal L}_{\rm eff}$ can be written in the form	
\beq
{\mathcal L}_{\rm eff}^{(2)}= \frac{1}{2} \int d^4 x \left(-i D^{\mu\nu} (x)\right) \Pi_{\mu\nu}^{(F^2)} (x)\,,
\label{10}
\eeq
where  $\Pi_{\mu\nu}(x) $ is the polarization operator:
\beq
\Pi_{\mu\nu} =  i \langle T\left\{ j_\mu (x)  j_\nu (0) \right\}\rangle\,,\qquad j_\mu = \bar \psi\gamma^\mu \psi\,.
 \eeq
The superscript $(F^2)$ in (\ref{10}) means that in the operator expansion for $\Pi_{\mu\nu}$  we are interested only in the single operator
$F_{\mu\nu} (0)F^{\mu\nu}(0)$:
\beq
\Pi_{\mu\nu}(x) = ... + C_{\mu\nu}(x)  F^2 (0) + ...\,.
\label{12}
\eeq
If we consider ${\mathcal L}_{\rm eff}$ in two loops, the coefficient $C_{\mu\nu}(x) $  must be calculated in the one-loop approximation. The expression for 
$C_{\mu\nu}(x) $ in this approximation is well defined -- it requires neither infrared nor ultraviolet regularization. In fact, $C_{\mu\nu}(x) $ is obtained by simple multiplication of the two electron propagators $S(0, x)$ and $S(x, 0)$. The quantity $\Pi_{\mu\nu}$	appearing	in (\ref{10})  has the form
const$\cdot F^2(0)\,x^{-2}$, and the factor $\log (M_0/\mu)$ (see (5)) arises from the integration over $d^4x$ that is performed at the very end. 
In fact, according to (\ref{10}) we have
\beq
{\mathcal L}_{\rm eff}^{(2)}(0) \sim \left(\int d^4 x \frac{1}{x^4} \right)F^2 (0) \sim \left(\log \frac{|x_{\rm min}|}{|x_{\rm max}|}\right) F^2 (0)\,.
\nonumber
\eeq

The construction of the operator product expansion for $\Pi_{\mu\nu}(x) $
 has been discussed repeatedly in the literature in
connection with the QCD sum rules [1]. This problem -- the
determination of the $F^2$ operator in $\Pi_{\mu\nu}$ -- has been discussed
in full detail in the review Ref. 3 (p. 609). It is instructive,
however, to go through this exercise again so that we can 
then stress those aspects that distinguish spinor electrodynamics from the scalar model and supersymmetric model.

First of all, for the external photon field we use the
Fock-Schwinger gauge [2,4,5]
\beq
x^\mu A_\mu (x) =0\,. 
\label{13}
\eeq
(A review of this technique is given in Ref. 3.) In this gauge 
the four-potential $A_\mu (x)$  is expressed in terms of the fields strength 
tensor $F_{\mu\nu} $:
\beq
A_\mu (x)= \frac{1}{2\cdot 0!}\, x^\rho F_{\rho\mu}(0) +\frac{1}{3\cdot 1!}\, x^\alpha x^\rho \left( \partial_\alpha F_{\rho\mu}(0) \right)+ ...
\label{14}
\eeq	
\begin{flushright}
\color{blue} 322
\end{flushright}
In fact, since we are interested only in the $F^2$ structure in the
effective Lagrangian, it is possible (and necessary) to confine ourselves to just the first term of the 
expansion in the right-hand side of (\ref{14}). This corresponds to a constant external electromagnetic field. Henceforth, terms with 
derivatives of $F$ will be consistently omitted. 

If we neglect the mass, the electron propagator will
have the form
\beqn
S(x, 0)&=& \left\langle x\left |\, \frac{1}{\hat{\mathcal P}} \, \right |0 \right\rangle =  -i\langle  T\{ T \psi(x)\bar\psi(0)\}\rangle
\nonumber
\\[2mm]
&=& \frac{1}{2\pi^2}\,\frac{\hat x}{x^4} -\frac{1}{8\pi^2} \,\frac{x^\alpha}{x^2}\, \tilde{F}_{\alpha\varphi} \gamma^\varphi\gamma^5 +...
\label{15}
\eeqn
where $\hat x=\gamma^\mu x_\mu$ and the ellipses denote operators that cannot give $F^2$  in ${\mathcal L}_{\rm eff}$  (e.g., 
$\partial_\alpha F_{\beta\gamma}$ or $F_{\mu\alpha} F^{\nu\alpha} - \frac 14 \delta^\nu_\mu F_{\alpha\beta}F^{\alpha\beta}$.

The absence in (\ref{15}) of a term of the form $(\hat x (\log x^2) F_{\alpha\beta}F^{\alpha\beta}$ has an independent theoretical explanation.
It follows from the general theorem [5] which states that $S(x, 0)$ has no logarithmic singularity in a self-dual external field. Dimensionally, generally speaking, a non-singular term $\hat x F_{\alpha\beta}F^{\alpha\beta}$ could arise. For the following analysis it is important that there should not in fact be such a term. We note that the method described, e.g., in Ref. 6 for constructing propagators in an external field fixes only the singular terms. Therefore, in order to elucidate whether or not a term $\hat xF^2$ is present in $S(x, 0)$ it is convenient to make direct use of the equation for the propagator $S(x, 0)$:

 \begin{figure}[h]
\centerline{\includegraphics[width=9.5cm]{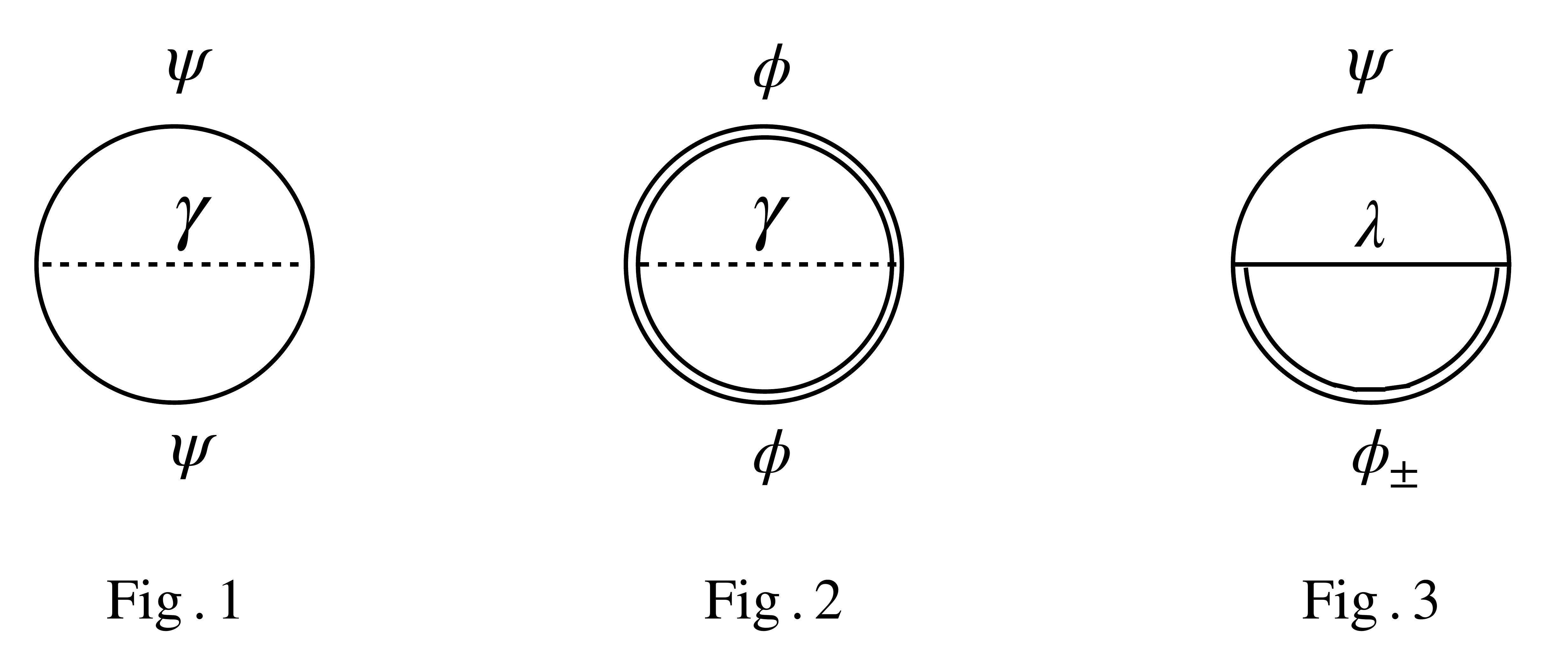}}
\label{khri} 
\end{figure}
\noindent
{\small {\em Figure captions}. Fig. 1: Too-loop graph for ${\mathcal L}_{\rm eff}$ in spinor electrodynamics, Sec. 3. The solid line is the electron propagator in the background field; Fig. 2: Too-loop graph for ${\mathcal L}_{\rm eff}$ in scalar electrodynamics (Sec. 4). The double solid line is the $\phi$ field propagator in the background field; Fig. 3: Additional contribution due to the photino exchange in supersymmetric electrodynamics (Sec. 5). The single and double solid lines 
denote the propagators of the spinor and scalar fields, respectively, in the external field.} 
\begin{flushright}
\color{blue} 323
\end{flushright}
\nopagebreak

\beq
i \gamma^\mu D_\mu S(x,0) = \delta^{(4)} (x)\,.
\label{16}
\eeq
If we substitute $A_\mu (x)$ in the form
\beq
A_\mu = \frac 12  x^\rho \, F_{\rho\mu}
\label{17}
\eeq
it is easily verified that there is no term $\hat x F^2$ in $S(x, 0)$. However, a nonsingular term $\sim  F^2$ appears and plays an important role in the propagator of a scalar particle. We shall postpone the relevant discussion to Sec. 4.

After these preliminary comments, it will not seem surprising to the reader that the subsequent calculation of ${\mathcal L}_{\rm eff}$  amounts to two or three simple algebraic operations. First of all,
\beq
\Pi_{\mu\nu}^{\,\,(F^2)} (x) = i {\rm Tr} \left\{\gamma_\mu S^{(F)} (x, 0)\,  \gamma_\nu S^{(F)}  (0, x)\right\}
=  \frac{-i}{192\pi^4}\, F^2(0) \,\frac{2x_\mu x_\nu + x^2 g_{\mu\nu}}{x^4}\,,
\eeq
where we have taken into account that
\beqn
&&S(x,0) = \gamma^0 S^\dagger (0,-x)\gamma^0;\nonumber\\[2mm]
&&\tilde F_{\alpha\beta} (0)\, \tilde F_{\gamma\delta}(0) \to -\frac{1}{12} \, F^2 (0)\,\big(g_{\alpha\gamma} g_{\beta\delta} - g_{\alpha\delta} g_{\beta\gamma} \big).
\nonumber
\eeqn
The result for $\Pi_{\mu\nu}^{\,\,(F^2)} $ turns out to be automatically transverse, as it should be for a $T$-product of conserved currents.

 It remains to take the last step. Substituting (18) and (9) into (10) and going over to Euclidean space $(x_0\to ix_4)$, we obtain
 \beq
 {\mathcal L}_{\rm eff}^{\,\, 2} = -\int _{|x|_{\rm min}}^{|x|_{\rm max}}\, \frac{d^4 x}{x^4}\, \frac{e_0^{\,^2}}{256\pi^6}\,F_{\mu\nu}(0) F^{\mu\nu}(0)
 =-\frac{e_0^{\,^2}}{2^7 \pi^4}\,\left(\log\frac{M_0}{\mu} \right)F^2 (0)\,,
 \label{19}
 \eeq
where $|x|_{\rm min}=M_0^{\,-1}$ and $|x|_{\rm max}=\mu^{\,-1}$.
In terms of $\beta^{(2)}$  (see (5))  the result (19) obviously reduces to
\beq
\beta^{(2)} = \frac{1}{2\pi^2} \qquad ( {\rm spinor \,\, electrodynamics} )\,,
\eeq
which 
coincides with the well known expression for the second coefficient of the Gell-Mann--Low function (see, e.g., Ref. 1). 
\begin{flushright}
\color{blue} 323
\end{flushright}

\section{Scalar electrodynamics}

The matter Lagrangian in the scalar case is $\Delta {\mathcal L} = \big(D_\mu\phi\big)^\dagger \big(D_\mu\phi\big)$, where the mass term has been 
omitted.
 The effective Lagrangian in the two-loop approximation is determined by the graph in Fig. 2.
 
Now, when we have formulated the main stages of the procedure for the example of spinor electrodynamics, the calculation of  $\beta^{(2)} $ in scalar electrodynamics proceeds considerably faster. We shall not dwell on those points that are the same in the two models.

First of all, the general expressions (10), (11) determining  ${\mathcal L}_{\rm eff}^{(2)}$ remain valid,\footnote{We note that the tadpole-type graphs generated by the contact term
$A_\mu A^\mu \phi^\dagger\phi $ does not give a contribution to the $\beta$ function and for this reason is not considered.}
the only difference being that the particle current now has the form
\beq
J_\mu = i\phi^\dagger \stackrel{\leftrightarrow}{D_\mu}\phi =i\big[ \phi^\dagger {D_\mu}\phi -\big( D_\mu \phi)^\dagger \phi\big]\,.
\eeq
Next, since a derivative appears in the definition of the current, to calculate $\Pi_{\mu\nu}$ it is necessary to know the function describing the propagation of a scalar particle from the point $y$ to the point $x$ and only after the differentiation can we set $y = 0$.\footnote{The Fock-Schwinger gauge $x_\mu A^\mu=0$ distinguishes the coordinate origin. Therefore, in gauge non-invariant quantities there is no translational invariance and $G(x,y)\neq G(x-y,0)$, where $G$ is the propagator of the scalar field.\normalsize \begin{flushright}
\color{blue} 323
\end{flushright}}

The Green function of the massless scalar field has the form
\beqn
G(x,y)&=& \left\langle x\left|\,\frac{1}{{\mathcal P}^2}\, \right| y\right\rangle=-i\langle T\{ \phi(x)\phi^\dagger (y)\rangle\nonumber\\[2mm]
&=& \frac{i}{4\pi^2}\,\frac{1}{(x-y)^2} + \frac{1}{8\pi^2}\,\frac{x^\mu y^\rho}{(x-y)^2}\, F_{\mu\rho}(0)\nonumber\\[2mm]
&=& \frac{i}{512\pi^2}\, (x-y)^2\, F^2(0) - \frac{i}{384\pi^2}\, \frac{x^2y^2 - (xy)^2}{(x-y)^2}\, F^2(0) + ...\,.
\eeqn
Here ... denotes terms with derivatives of $F_{\mu\nu}$ and terms of the type
$$F_{\mu\alpha} F^{\nu\alpha} - \frac 14 \delta^\nu_\mu \big( F_{\alpha\beta}F^{\alpha\beta}\big)\,.$$ The last term in the right-hand side does not give a contribution 
to ${\mathcal L}_{\rm eff}^{(2)}$ since the coefficient in it is proportional to $y^2$. The most important term in (22) is the third, the appearance of which was for us a surprise. In fact, it appears with a coefficient that is nonsingular as $x \to y$, and so is not fixed in the framework of the standard procedure of expanding in the momenta (see Ref. 6). Thus, in this aspect the situation differs radically from that which we obtained for the spinor propagator. In the case of the spinor Green function all the terms necessary for the calculation of ${\mathcal L}_{\rm eff}^{(2)}$ had singular coefficients and were determined in the framework of the standard OPE procedure (i.e., expansion in singularities).

The fact that $( x - y)^2 F^2$ is present in $G(x,y)$ is easily checked using the equation of motion:
\beq
- D_\mu^2\, G(x,y) =\delta^{(4)} (x-y)\,.
\eeq
This exercise becomes especially simple if we set $y = 0$. Then, from general arguments,
\beq
G(x, 0)= \frac{i}{4\pi^2 x^2}+C x^2 F^2 (0)+O(F^3)\,.
\nonumber
\eeq
Here $C$ us a certain constant (in $G(x,0) $ there can be no term linear in $F$). Then from (23)
\beq
-D^2 G(x,0) = \left[-\left(\frac{\pt}{\pt x^\mu}\right)^2+\big(A_\mu(x)\big)^2 +2i A^\mu\pt_\mu
\right] G(x,0) = \delta^{(4)} (x)
\nonumber
\eeq
where we have made use of the fact that, by virtue of (17),
$\pt^\mu  A_\mu (x) = 0$. Next, $A_\mu \pt^\mu G(x, 0)$ can be omitted
because of the absence in $G(x, 0)$  of a term linear in $F$. Finally,
the relation
\beq
\left\{\left[ -\left(\frac{\pt}{\pt x} \right)^2+ \big(A_\mu (x)\big)^2 \right] G(x,0)\right\}_{F^2} = 0\,, \qquad \left( A_\mu^{\,2} = \frac{1}{16} x^2 F^2(0)\right)
\nonumber
\eeq
makes it possible to determine the constant C:
\beq
C=\frac{i}{512\pi^2}\,.
\nonumber
\eeq
It is usually assumed that terms nonsingular in $x$ (in
momentum space they have the form of a $\delta$ function and
derivatives, e.g., $x^2 \leftrightarrow [(\pt/\pt q_\mu )^2 \delta^{(4)} (q)]$  are connected not
with small but with large distances. In the present case we
shall see that this is not so. The term $(x- y)^2 F^2 (0)$, like
other terms in (22), comes from short distances, and cannot
be dropped from (22) without violating the equations of
motion.
\begin{flushright}
\color{blue} 324
\end{flushright}

\newpage

Indeed, it is clear that this contribution is important for
calculating the part
singular in $x$ 
in $\Pi_{\mu\nu}(x)$. The reason is
that in $\Pi_{\mu\nu}(x)$ the part that is polynomial in $x$ is multiplied
by the singular propagator of the free scalar field (more precisely,
by its derivative). As a result, the product is singular.
Although, in principle, the procedure for constructing $\Pi_{\mu\nu}^{\,\,F^2}(x)$
is the same as for the spinor case analyzed above, there is a
slight technical complication associated with the presence of
the covariant derivative in the definition (21) of the current.
If $ y \to 0$, then$  D_{\mu\,y}$  can be assumed to coincide with the
ordinary derivative $\pt/\pt y^\mu$. In the one-loop approximation
$\left.\Pi_{\mu\nu}(x,y)\right|_{y\to 0}$ has the form
\beq
\Pi_{\mu\nu}(x,y) = 2i \Big\{
D_\mu G(x,y) \big(\pt /\pt y^\nu\big) G(y,x) - D_\mu G(x,y) \big(\stackrel{\leftarrow}{\pt} /\pt y^\nu\big) G(y,x)\Big\}\,.
\eeq
If we make use of the expression (17) for $A_\mu(x)$, then in
$\Pi_{\mu\nu}^{\,(F^2)}(x,y\to 0)$ two types of contribution arise:
\beq
\Pi_{\mu\nu}^{\,( F^2)}= \Pi_{\mu\nu}^{\,\,(1)}+ \Pi_{\mu\nu}^{\,\,(2)}\,,
\nonumber
\eeq
where
\beqn
\Pi_{\mu\nu}^{\,\,(1)} &=&x^\rho F_{\rho\mu}(0)\left[ G(x,y) \frac{\pt}{\pt y^\nu} G(y,x) - G(y,x) \frac{\pt}{\pt y^\nu} G(x,y)
\right]_{y=0}
\nonumber\\[2mm]
& =& -\frac{i}{192\pi^4x^4}\big( x^2 g_{\mu\nu} -x_\mu x_\nu\big) F^2 (0)\,;
\nonumber\\[2mm]
\Pi_{\mu\nu}^{\,\,(2)} &=& 2i\left[ \frac{\pt}{\pt x^\mu} G(x,y) \frac{\pt}{\pt y^\nu} G(y,x) - \left(\frac{\pt}{\pt x^\mu} \frac{\pt}{\pt y^\nu} G(x,y)\right)  G(y,x)
\right]_{y=0}
\nonumber\\[2mm]
& =& -\frac{i}{64\pi^4x^4}\big(x_\mu x_\nu\big) F^2 (0)\,,
\eeqn
where we have used the explicit expression (22) for the propagator.
We stress that $\Pi_{\mu\nu}^{\,\,(2)}$  owes its origin entirely to the
nonsingular term in (22). If we had not taken the latter into
account, we would have obtained a non-transversal  result for
$\Pi_{\mu\nu}$. 

Collecting $\Pi_{\mu\nu}^{\,\,(1)} $  and $\Pi_{\mu\nu}^{\,\,(2)} $, we arrive at 
\beq
\Pi_{\mu\nu}^{\,( F^2)} = -\frac{i}{192\pi^4x^4}\big( x^2 g_{\mu\nu} +2x_\mu x_\nu\big) F^2 (0)\,,
\eeq
which satisfies the transversality condition $(\pt/\pt x_\mu ) \Pi_{\mu\nu}=0$
and coincides numerica1ly with the result (18). Therefore, without repeating the subsequent calculations, we conclude that\nopagebreak\begin{flushright}
\color{blue} 324
\end{flushright}
\newpage
\beq
\beta^{(2)} = \frac{1}{2\pi^2} \qquad (\rm scalar\,\, electrodynamics)\,,
\eeq
in complete agreement with the literature (see, e.g., Ref. 1).

\section{Supersymmetric electrodynamics}

In superfield notation the action has the form
\beq
S_{\rm SUSY\, QED}= \frac{1}{4e_0^2} \int d^2\theta\, d^4 x\, W^2
+ \frac 14  \int d^2\theta\, d^2\bar\theta\, d^4 x\left( \bar Te^VT + \bar Se^{-V} S\right)\,,
\eeq
where $T$ and $S$ are two left-handed chiral superfields with
opposite charges, $V$ is a real superfield, incorporating the
field of the photon and photino, and $W$ is the stress superfield.
 In components, the SQED Lagrangian is
\beqn
{\mathcal L} &=& -\frac{1}{4e_0^2} \, F_{\mu\nu}^{\,\,2} +\frac{i}{e_0^2}\, \bar\lambda \sigma^\mu\pt_\mu\lambda 
+\sum_{q=\pm}\left[\big(D_\mu \phi_q\big)^\dagger \big(D^\mu \phi_q\big) +\bar\psi_q i\sigma^\mu D_\mu\psi_q\right.
\nonumber\\[2mm]
&-&\left. i\sqrt{2} \,\phi_q^\dagger\big(\lambda\psi_q\big) + i\sqrt{2} \,\big(\bar\psi_q\bar\lambda\big)\phi_q  \right]- e_0^2 \big(\phi_+^\dagger\phi_+
-\phi_-^\dagger\phi_- \big)^2\,.
\eeqn
In the language of components the matter sector inclues
two Weyl spinors $\psi^\alpha_q, \,\, (\alpha= 1,2)$  wth charges $q =\pm$
which can be combined into a single Dirac spinor $\Psi$ 
describing the electron. To each Weyl spinor corresponds its 
complex scalar field $\phi_q$. Both the electron and the scalar fields
have the usual gauge coupling with the photon
$(D_\mu  =\pt_\mu \mp iqA_\mu)$.

The gauge sector contains not only the photon but also
the photino (charge 0), described by the Weyl spinor $\lambda^\alpha$ where $\alpha = 1,2$. 
The self-interaction of the scalar field, (the square
of the $D$ term) has no effect in the calculation of ${\mathcal L}_{\rm eff}^{(2)}$
and appears only m higher orders. 

The calculation is conveniently performed by going over 
to the Dirac spinor
\beq
\Psi= \left(
\begin{array}{c}
\psi_+
\\[1mm]
\bar\psi_-
\end{array}
\right).
\nonumber
\eeq
In this notation the vertex of the interaction of $\lambda$  with $\Psi$ takes the form
\begin{flushright}
\color{blue} 324
\end{flushright}

\newpage
\beq
{\mathcal L}_{\rm int} =-\sqrt{2}\, i\, \left[ \phi_+^\dagger\left(\bar\lambda \frac{1+\gamma^5}{2}\Psi\right) 
+\phi_-^\dagger\left(\bar\lambda \frac{1-\gamma^5}{2}\Psi\right) 
\right]+{\rm H.c.}\,.
\eeq
Thus, in the calculation of ${\mathcal L}_{\rm eff}^{(2)}$ in supersymmetric electrodynamics
it is necessary to take into account all three diagrams depicted in Figs. 1, 2, and 3. 
(In the diagrams of Figs. 2 and 3 scalar  
particles of two types $q = \pm$  are propagating.) We  have
found the diagrams of Figs. 1 and 2 in the preceding sections.

The only additional contribution is connected with the graph 
in Fig. 3.

As in the case of photon exchange, we take into account
the fact that the photino does not interact directly with the external 
field. Then ${\mathcal L}_{\rm eff}^{(2)}$(Fig.3) is represented in the form
\beq
{\mathcal L}_{\rm eff}^{(2)}(\,{\rm Fig.\,3})=-
2i \int _{|x|_{\rm min}}^{|x|_{\rm max}}\, d^4 x\, \left\langle T \big\{ \lambda_\beta (0)\bar\lambda(x) \big\}\right \rangle\, \Pi_{\alpha\beta}^{\,\,(F^2)} (x)
\eeq
where the polarization operator $\Pi_{\alpha\beta}$  is defined as
\beq
\Pi_{\alpha\beta }(x) = \left\langle T \big\{\bar\phi (x) \Psi_\alpha (x), \phi(0) \bar\Psi_\beta (0)\big\} \right\rangle =i^2 G(0,x) S_{\alpha\beta}(x,0)\,.
\eeq

In this expression for $\Pi_{\alpha\beta }$  we have already taken into account 
both types scalar. We recall that the photino propagator is free
and has the form
\beq
\left\langle T \big\{ \lambda_\beta (0)\bar\lambda(x) \big\}\right \rangle=-\frac{ie_0^2}{2\pi^2}\,\frac{\hat x_{\alpha\beta}}{x^4}
\,.
\eeq
Since the expressions for the propagators $S(x, 0)$ and $G(0,x)$
are known (see (I15) and (22)), the calculation of $\Pi_{\alpha\beta}^{\,\,(F^2)} (x)$
proceeds trivially:
\beq
\Pi_{\alpha\beta}^{\,\,(F^2)} =-\frac{i}{1024\pi^4}\,\frac{\hat x_{\alpha\beta}}{x^2}\,F^2(0)\,.
\eeq
We note that only the $F^2$  part in $G(0, x)$ and the free term in $S(x,0)$
have cooperated here. Substituting (34) and (33) into (31)
we obtain
\beq
{\mathcal L}_{\rm eff}^{(2)}(\,{\rm Fig.\,3})= \frac{e_0^2}{2^7\pi^4}\left(\log\frac{M_0}{\mu}\right) F^2(0)\,.
\eeq
In terms of $\beta^{(2)}$ (see ( 5)) the result (35) obviously reduces to
\beq
\beta^{(2)}\,({\rm Fig.\,3})= -\frac{1}{2\pi^2}\,.
\eeq
Adding now the electron loop and the two loops with the
scalars (Fig. 2), we arrive 
\begin{flushright}
\color{blue} 325
\end{flushright}

\newpage
at the conclusion that in
supersymtnetric electrodynamics
\beq
\beta^{(2)} =\frac{1}{\pi^2}\qquad  \, {\rm (SUSY\,\, QED)}\,.
\eeq

\section{ Conclusion}

The question of the calculation of the second coefficient
of the Gell-Mann-Low function arose in connection with
the fact that the ratio $\beta^{(2)}/\beta^{(1)}$  in supersymmetric electrodynamics
has been obtained recently by a completely different
method [7] and it was desirable to compare the prediction of
Ref. 7 with direct calculations. Since, unfortunately, we did
not succeed in finding standard calculations of $\big(\beta^{(2)}/\beta^{(1)}\big))_{\rm SUSY\, QED} $ 
 in the literature, it was necessary to devise a method that
would make it possible to calculate the graphs for $\beta^{(2)}$  within
a reasonable interval of time. In this way, we have obtained
\beq
\beta (\alpha)_{\rm SUSY\, QED} =\frac{\alpha^2}{\pi} \left(1 + \frac{\alpha}{\pi} + ...
\right),
\eeq
which is in agreement with the prediction of Ref. 7, according
to which
\beq
\beta (\alpha)_{\rm SUSY\, QED} =\frac{\alpha^2}{\pi} \big(1 + \gamma_m\big) =  \frac{\alpha^2}{\pi} \left(1 + \frac{\alpha}{\pi} + ...
\right),
\nonumber
\eeq
where $\gamma_m$ is the anomalous mass dimension. The above calculation of $\beta^{(2)}$,
in our opinion, convincingly demonstrates
the effectiveness of the method of calculating the coefficients f$\beta^{(i)}$ by means of the operator-product-expansion method.
The two-loop calculation is maximally simplified, since it
uses prepared blocks, i.e., a simple multiplication of known
propagators takes place. In addition, it is not necessary to
display particularly high accuracy in the determination of
the regularization procedure.

A new aspect for us was the necessity of taking account
of terms nonsingular in $x$ in the propagator of the scalar
field. Formally, this corresponds to the region of virtual momenta
of the order of the momenta of the external field, although
the result for the polarization operator corresponds
to the normal OPE. Further discussion of this question -- the
relationship between the Wilson operator product expansion
and the analysis performed in scalar and supersymmetric
electrodynamics -- will be given in a separate publication [8].
\begin{flushright}
{\em Translated from Russian by P.~J.~Shepherd}\\[1mm]
\color{blue} 325
\end{flushright}

\vspace{-1mm}

\subsection*{Addendum to section 4, 1992, from [9]}

An important aspect aspect to be elucidated now is the interpretation of the non-singular terms in the
operator product expansion discussed in Sec. 4 (see Eq. (23) and below). Can one find a place for such terms within the consistent
procedure of separation of short- and large-distance contributions?

To answer this question let us turn to the consideration of $\Pi_{\mu\nu} $ in scalar QED. The
puzzling  term $x^2\,F^2$  in $G(x,0)$ below Eq.~(23)  corresponds to the diagram presented in Fig. 4$a$. Since the
propagation function in the lower part of the graph is proportional to $\delta^{\prime\prime}(p)$, following the
ideology of OPE we must actually cut the lower line and then the upper line shrinks to a point
(Fig.~4$b$). Thus, the calculation of Fig.~4$a$ is a two-step process. First, within. the standard
OPE approach, we calculate the coefficient in front of the operator
$(D_\mu\phi)^\dagger )( D^\mu \phi )$
(see Fig. 4$b$) -- this coefficient is determined entirely by short distances -- and, then, the conversion of
$(D_\mu\phi)^\dagger )( D^\mu \phi )$  into $F_{\mu\nu}^{\,\, 2}$ which can be ascribed to large distances. 
In other words, the second
step is obviously the calculation of a photonic matrix element of $(D_\mu\phi)^\dagger )( D^\mu \phi )$.

\begin{figure}[h]
\centerline{\includegraphics[width=6cm]{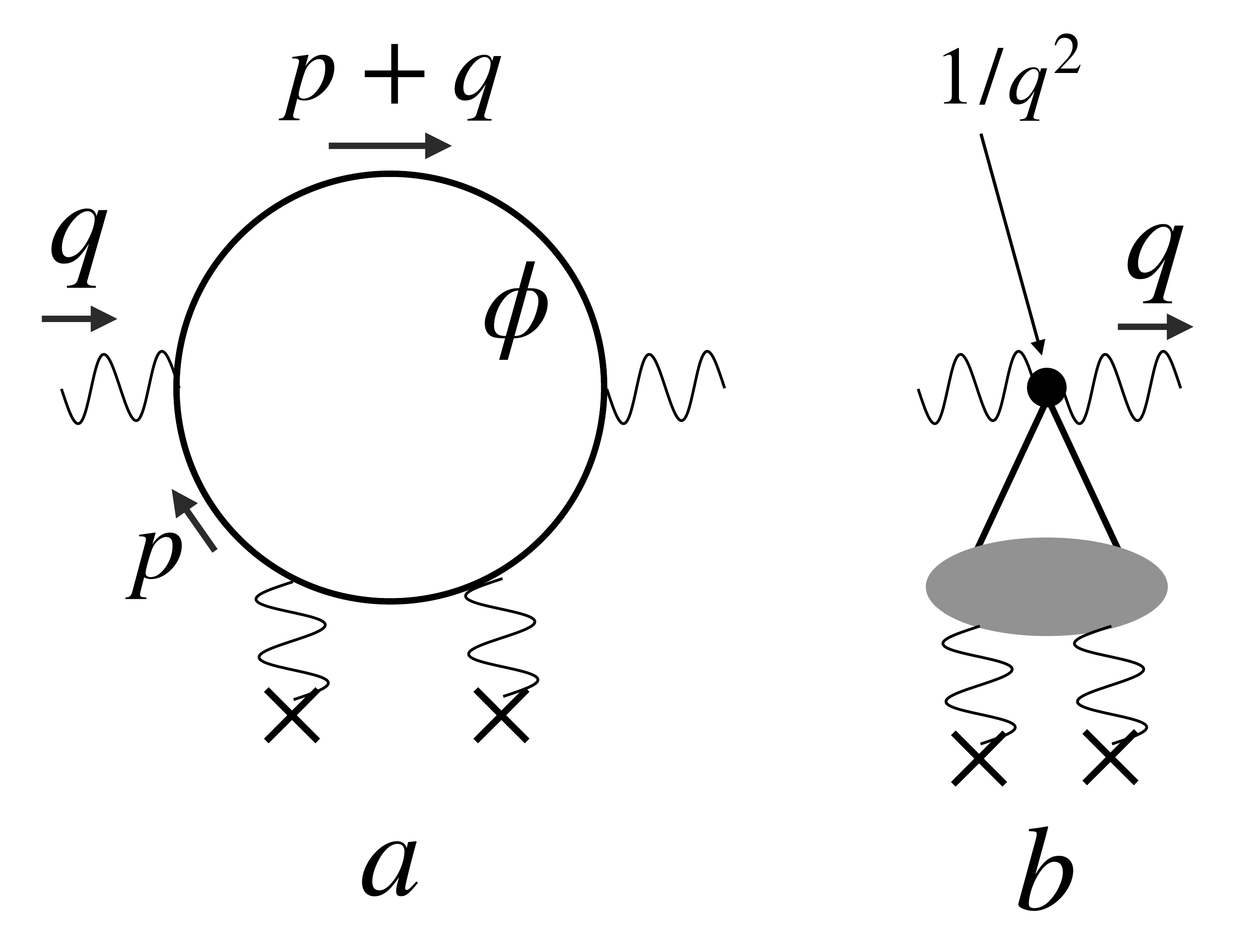}}
\label{khri} 
\end{figure}
\noindent
{ Fig. 4.\small {\em Figure caption}. Operator product expansion associated with the non-singular part in Eq. (22) (see Sec. 3) and the emergence of the condensate (39), (40).}

\vspace{1mm}

The last term in eq. (22) lead to the ``normal" OPE for  $\Pi_{\mu\nu}^{\,\, (F^2)}$ with the coefficient
function determined entirely by short distances.

Formally, the operator $\int d^4 x (D_\mu\phi)^\dagger )( D^\mu \phi )$  vanishes because of the equations of motions.
One can readily convince oneself, however, that in the external gauge field there is an
``anomalous" relation
\beq
\left\langle \big(D_\mu\phi\big)^\dagger )\big( D^\mu \phi \big)\right\rangle 
\eeq
stemming from the first term in Eq. (22).

The full Green function $G(x, y)$ certainly satisfies the equation of motion $$-D^2 G(x,y) =\delta^4 (x-y)\,.$$
However, in calculating  $\Pi_{\mu\nu}^{\,\, (F^2)}$ we split it in two pieces: the piece singular in $x$ is
used for determination of the coefficient function while the piece regular in $x$, $G^{\rm reg}(x, y )$, is
interpreted as a matrix element. Then $ \langle  (D_\mu\phi)^\dagger )( D^\mu \phi )\rangle \neq 0$. More specifically, Eq. (22)
implies
\beq
\left\langle \big(D_\mu\phi\big)^\dagger )\big( D^\mu \phi \big)\right\rangle = \lim_{x\to 0}\Big(-i D_\mu D^\mu \, G^{\rm reg} (x,0)\Big)
=\frac{1}{64\pi^2}\,F_{\alpha\beta} F^{\alpha\beta}\,.
\eeq
Now we are finally able to explain the difference between our first result for $\Pi_{\mu\nu}$  quoted in the second line in (25) 
and the complete answer presented in Eq. (26). The former expression  has been obtained
by substituting in $\Pi_{\mu\nu}$ the singular part of $G(x, y)$. Thus, it corresponds, in the
language of OPE, to the genuine contribution of the operator $F^2$. 

One should not forget, however, another dim = 4 operator, $(D_\mu\phi)^\dagger )( D^\mu \phi )$, whose matrix
element in the background electromagnetic field reduces to the same structure $F^2$ (see Eq. (40)). 
The coefficient of $(D_\mu\phi)^\dagger )( D^\mu \phi )$ can be trivially extracted from the diagrams in Fig. 5.
With no effort we arrive at
\beq
\Pi_{\mu\nu} (q) =-2 \left(\frac{g_{\mu\nu}}{q^2}  -2\,\frac{q_\mu q_\nu}{q^4}
\right) \left\langle \big(D_\mu\phi\big)^\dagger )\big( D^\mu \phi \big)\right\rangle .
\eeq
Invoking Eq. (D.2) in Ref. [3] for the Fourier transformation and Eq. (40) we then find that
the contribution of Fig. 4$b$ is equal to
\beq
\frac{-i}{64\pi^2}\,\frac{x_\mu x_\nu}{x^4}\,F^2(0)\,,
\eeq
precisely the difference between the expression in the second line of  (25) and Eq. (26).

\vspace{0.5cm}

{\bf \large 
References}

\vspace{0.3cm}

{\small

[1]
M. A. Shifman, A. I. Vainshtein, and V.I. Zakharov, Nucl. Phys. {\bf B147},
385 (1979).

[2] J. Schwinger, {\sl Particles, Sources and Fields}, 
Vols. 1 and 2, (Addison-Wesley, New York, 1963); [2-nd Edition:  CRC Press, 1998].

[3]
V. Novikov, M. Shifman, A. Vainshtein and V. Zakharov, 
{\em Calculations in External Fields in Quantum Chromodynamics. Technical Review,}
Fortsch. Phys. \textbf{32}, 585 (1984), 
see also  [9], pp. 236-268.
  
[4]
V. A. Fock, {\em Proper time in classical and quantum mechanics}
Phys. Z. Sowjet\-union {\bf12}, 404-425, (1937); {\sl Works on Quantum Field
Theory}, (Leningrad University Press, 1957), p. 150.

[5]
V. A. Fateev, A. S. Schwarz, Yu.S. Tyupkin, {\em On particle-like solutions in the presence of fermions}, 
Preprint  FIAN- 155,  1976; C.
Cronstr\"om, Phys. Lett. {\bf 90B}, 267 (1980); M.~S. Dubovikov and A. V. Smilga, Nucl. Phys. {\bf B185}, 109 (1981).

[6] E. V. Shuryak and A. I. Vainshtein, Nucl. Phys. {\bf B201}, 141 (1982).
 
 [7]
 A. I. Vainshtein, V.I. Zakharov, and M.A. Shifman, 
JETP Lett. 42, 224 (1985), M.~A.~Shifman, A.~I.~Vainshtein and V.~I.~Zakharov,
{\em Exact Gell-Mann--Low Function in  Supersymmetric  Electrodynamics},''
Phys. Lett. B \textbf{166}, 334 (1986).
   
[8]
M.~A.~Shifman and A.~I.~Vainshtein,
{\em Solution of the Anomaly Puzzle in SUSY Gauge Theories and the Wilson Operator Expansion,}
  Nucl.\ Phys.\ B {\bf 277}, 456 (1986).
  
 [9] M. Shifman, {\em Vacuum Structure and QCD Sum Rules},  in {\sl Current Physics Sources and Comments}, (North-Holland, 1992), Vol. 10, p. 267-268.

 }

\end{document}